# On conformal lenses

*Huanyang Chen\*, Yadong Xu, and Hui Li*
*School of Physical Science and Technology, Soochow University, Suzhou, Jiangsu 215006, China*

**Abstract:** Plane mirror can make one object into two for observers *on the object's side*. Yet, there seems no way to achieve the same effect for observers *from all directions*. In this letter, we will design a new class of gradient index lenses from multivalued optical conformal mapping. We shall call them the conformal lenses. Such lenses can transform one source into two (or even many) omnidirectionally. Like the overlapped illusion optics does, they can even transform multiple sources into one. Rather than using negative index materials, implementation here only needs isotropic positive index materials like other gradient index lenses. One obvious drawback however, is that they have singular permittivity values which restrict them to functioning at one single frequency. This however, needs not be the case when applying transmutation methods, which enable the lenses to work in a broadband frequency range.

Gradient index lenses [1], such as the Maxwell fish-eye lens [2], the Luneburg lens [3], and the Eaton lens [4] have recently drawn much attention because of their utility for designing invisibility cloaks [5], omnidirectional retroreflectors [6] and various imaging functionalities [7,8,9,10,11,12] including the perfect imaging [13,14,15]. The conformal lenses here proposed are based on the optical conformal mapping [5], which shares the same spirit with transformation optics[16]. Transformation optics [5,16] offers us a versatile tool to control the electromagnetic field [17,18], in designing invisibility cloaks [19,20,21,22,23,24,25,26] and other novel devices (see a recent review [18]). Generally, materials designed by transformation optics are inhomogeneous and anisotropic. In particular, such anisotropic materials are often of magnetic response, leading to challenges in implementation. Optical conformal mapping [5] however, can be applied to obtain isotropic and non-magnetic response materials, thereby facilitating the implementation. There have been some optical conformal devices proposed, such as beam collimator [27,28,29] and beam bend [30].

Among different kinds of transformation optics, illusion optics [31] has been brought forth as a new concept to transform one object into another. Overlapped illusion optics [32] goes further to transform multiple objects into one, reversing the mirror effect, where with a plane mirror, one object can be transformed into two for observers *on the object's side*. In both cases, negative index materials are used. Now, another question comes out - is there a device to transform one object into two (or even many) for observers *from all directions*? At the first thought, it seems impossible. The reason is for passive image objects, there should be multiple scattering effects, which cannot be obtained from only one single object. However, we need not necessarily choose passive ones. When it comes to active image objects, since there is no such effect, it is then possible for one object to become two. In this letter, we will first introduce a simple multivalued conformal mapping. The conformal lenses devised here can not only transform one source into two, but also perform the inverse. What's more, the materials needed are isotropic just as that of the Eaton lens. Negative index materials [32] are hence not necessary now. Using the transmutation methods [33, 34], we can even remove the singular values of the material parameters, thus enabling the design of the lens for broadband frequencies. A group of multivalued conformal mappings will then be proposed for the lenses to transform one source into many, and vice versa.

---

\* chy@suda.edu.cn

We will start the story from the conformal mapping:

$$w^2 = z^2 - 1, \qquad (1)$$

which maps $w$-space into $z$-space (two values of $w$ are mapped into two values of $z$).

From the theory of optical conformal mapping [5], we should have the relationship of the refractive indexes in $w$-space ($n_w$) and $z$-space ($n_z$),

$$n_z = n_w | \frac{dw}{dz} |, \qquad (2)$$

by keeping the optical path unchanged during the mapping. In this case, if we suppose $n_w = 1$ (vacuum in $w$-space), the refractive index in $z$-space should be,

$$n_z = \frac{|z|}{\sqrt{|z^2-1|}} = \sqrt{\frac{x^2+y^2}{\sqrt{(x^2-y^2-1)^2+4x^2y^2}}}. \qquad (3)$$

Writing in bipolar coordinate, we have,

$$n_z = \sqrt{\frac{\cosh \tau + \cos \sigma}{2}}, \qquad (4)$$

where $x = \frac{\sinh \tau}{\cosh \tau - \cos \sigma}$ and $y = \frac{\sin \sigma}{\cosh \tau - \cos \sigma}$. From Eqs. (3) and (4), it's clear that three singular points can be found, i.e., when $z=0$, $n_z=0$, when $z=\pm 1$, $n_z=+\infty$. There are two kinds of special mappings here. One is that $z=0$ is mapped to $w=\pm i$ (one point mapped to two points). The other is that $z=\pm 1$ are mapped to $w=0$ (two points mapped to one point). These two properties would be used to demonstrate the following important functionalities, i.e., transforming one source into two, and two into one. The transformation media described in Eqs. (3) and (4) are called the conformal lenses, which can work for active sources in both wave optics realm and geometric optics limit.

It is important to note, optical conformal mapping usually requires materials to be filled in the whole space. However, for the mapping in Eq. (1), when $z \to \infty$, $w \to z$, we can introduce a cut-off radius $r_c$ so that,

$$n_z = \begin{cases} 1, & |z| > r_c, \\ \dfrac{|z|}{\sqrt{|z^2-1|}}, & |z| \le r_c, \end{cases} \quad (5)$$

indicating that the lens here is of finite size, with almost the same properties reserved as the full one in Eqs. (3) and (4) if $r_c$ is large enough. We will use this finite-size lens to demonstrate the intriguing wave functionality by performing numerical simulations below.

We will only consider the transverse electric (TE) polarized waves (i.e., $\varepsilon_z = n_z^2$, $\mu = 1$) and the line current sources. Here we assign $r_c = 5$ and the wavelength $\lambda = 2$ (arbitrary units). Figure 1a shows the electric field distribution when there is one current source (with a 1 A current) located at the position $z = 0$ in the conformal lens described by Eq. (5). The far field distribution (in the region $r \ge r_c$) appears to be that of two line current sources (each with a 1/2 A current) located at $w = \pm i$ in vacuum, which is illustrated in Figure 1b. In other words, one source is transformed into two. Figure 1c shows the electric field distribution when there are two line current sources located at the positions $z = \pm 1$ in the lens. Each one carries 1 A current. The far field distribution appears to be that of only one line current source with a 2 A current in vacuum, which is illustrated in Figure 1d. In other words, two sources are transformed into one.

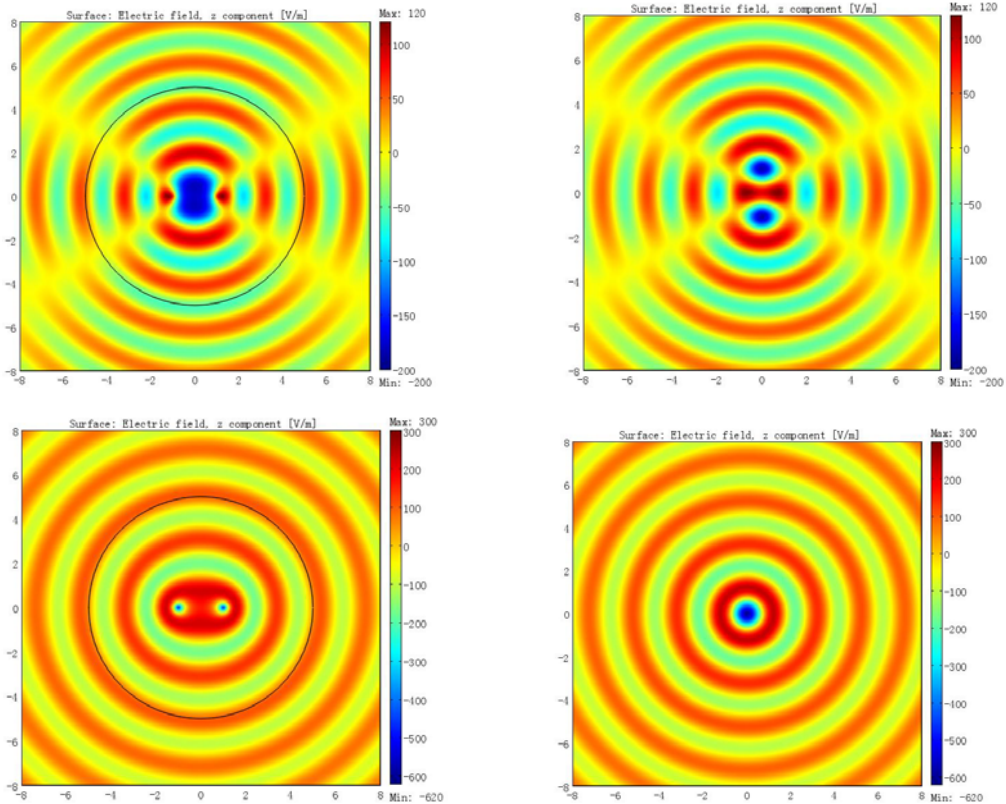

**Figure 1 | The electric field distribution for a,** one line current source with a 1 A current in the conformal lens described by Eq. (5); **b,** two line current sources (each with a 1⁄2 A current) in vacuum; **c,** two line current sources (each with a 1 A current) in the lens; **d,** one line current source with a 2 A current in vacuum.

As described previously, there are singular values among the permittivity distribution, the lens can only work at one single frequency and is very difficult to implement. However, we find that near the singular points,

$$n_z \propto \begin{cases} r_{+1}^{-\frac{1}{2}} = [\sqrt{(x-1)^2 + y^2}]^{-\frac{1}{2}}, & z \to +1, \\ r = \sqrt{x^2 + y^2}, & z \to 0, \\ r_{-1}^{-\frac{1}{2}} = [\sqrt{(x+1)^2 + y^2}]^{-\frac{1}{2}}, & z \to -1, \end{cases} \quad (6)$$

whose singular values can then be transmuted using coordinate transformation [33]. For example, we can perform the following transformation ($(r, \theta, z) \Leftrightarrow (R, \Theta, Z)$),

$$R = \begin{cases} \dfrac{r^2}{r_0}, & 0 \leq r \leq r_0, \\ r, & r > r_0, \end{cases} \quad \Theta = \theta, \quad Z = z, \quad (7)$$

to obtain the material parameters,

$$\mu_r = 2, \quad \mu_\theta = \frac{1}{2}, \quad \varepsilon_z = \frac{1}{2} \frac{b^2}{\sqrt{(x^2 - y^2 - 1)^2 + 4x^2 y^2}}, \quad (8)$$

inside the region $0 \leq r \leq r_0$. During the simulation, we set $r_0 = 0.5$. For the other two singular points, it is more convenient to perform the transmutation in bipolar coordinate ($(\tau, \sigma, z) \Leftrightarrow (T, \Sigma, Z)$),

$$T = \begin{cases} \dfrac{1}{2}(\tau + \tau_0), & \tau \geq \tau_0, \\ \tau, & -\tau_0 < \tau < \tau_0, \\ \dfrac{1}{2}(\tau - \tau_0), & \tau \leq -\tau_0, \end{cases} \quad \Sigma = \sigma, \quad Z = z. \quad (9)$$

The transformed parameters are,

$$\mu_\tau = \frac{1}{2}, \quad \mu_\sigma = 2, \quad \varepsilon_z = \frac{(\cosh\frac{1}{2}(\tau + \tau_0) - \cos\sigma)^2}{(\cosh\tau - \cos\sigma)^2}(\cosh\tau + \cos\sigma), \quad (10)$$

for $\tau \geq \tau_0$ (or inside the region $(x - \coth\tau_0)^2 + y^2 \leq \operatorname{csch}^2\tau_0$) and,

$$\mu_\tau = \frac{1}{2}, \ \mu_\sigma = 2, \ \varepsilon_z = \frac{(\cosh\frac{1}{2}(\tau-\tau_0) - \cos\sigma)^2}{(\cosh\tau - \cos\sigma)^2}(\cosh\tau + \cos\sigma), \quad (11)$$

for $\tau \leq -\tau_0$ (or inside the region $(x + \coth\tau_0)^2 + y^2 \leq \operatorname{csch}^2\tau_0$). To avoid overlapping with the region $0 \leq r \leq r_0$, we assign $\tau_0 = 1.2$ for instance during the simulation. Figure 2 shows the distribution of $\varepsilon_z$, which has no singular value now, thus enabling the lens to operate in a broad band of the spectrum if it is embedded in a dielectric with a suitable permittivity value.

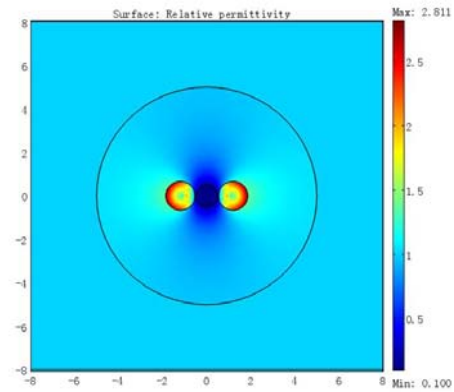

**Figure 2 | The permittivity distribution for transmuted conformal lens.**

Figure 3a shows the electric field distribution when there is only one line current source (with a 1 A current) located at the positions $z = 0$ in the transmuted conformal lens. The far field distribution appears to be that of two line current sources (each with a 1/2 A current) located at $w = \pm i$ in vacuum (see also in Figure 1b). Figure 3b shows the electric field distribution when there are two line current sources (each with a 1 A current) located at the positions $z = \pm 1$ in the transmuted lens. The far field distribution appears to be that of only one line current source with a 2 A current in vacuum (see also in Figure 1d).

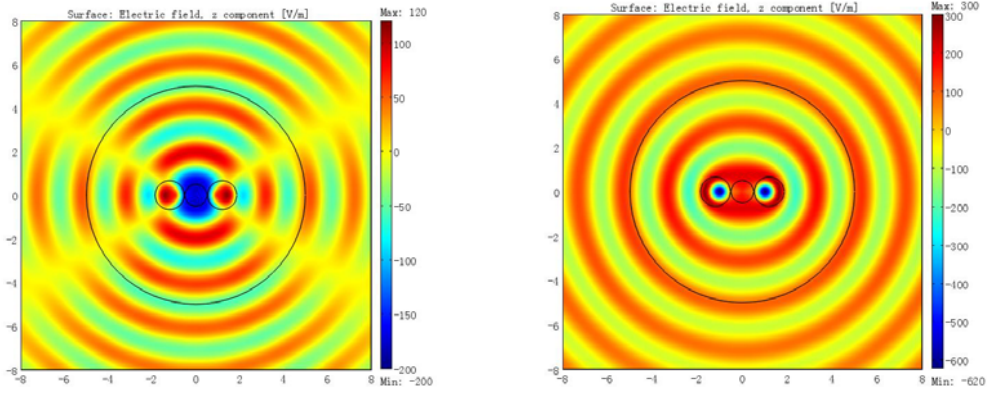

**Figure 3 | The electric field distribution for a,** one line current source with a 1 A current in the transmuted conformal lens; **b,** two line current sources (each with a 1 A current) in the transmuted lens.

Above we discuss how to transform one source into two and two into one. Now we will consider a more general conformal mapping:

$$w^N = z^N - 1, \qquad (12)$$

which maps $w$-space into $z$-space ($N$ values of $w$ are mapped into $N$ values of $z$).

The transformed refractive index in $z$-space is

$$n_z = \frac{|z|^{N-1}}{|z^N - 1|^{1-\frac{1}{N}}}, \qquad (13)$$

if the refractive index in $w$-space is unity. There are $N+1$ singular points, including $z=0$ (where $n_z = 0$) and $z = e^{2i\pi \frac{j}{N}}$ ($j = 0, 1, ..., N-1$) (where $n_z = +\infty$). Similar transmutation methods can be applied for finite values of the electromagnetic parameters. There are two kinds of special mappings. One is that $z = 0$ is mapped to $N$ points ($w = e^{i\pi \frac{2j+1}{N}}$ ($j = 0, 1, ..., N-1$)). The other is that $N$ singular points ($z = e^{2i\pi \frac{j}{N}}$ ($j = 0, 1, ..., N-1$)) are mapped to $w = 0$. Such intriguing properties enable us to transform one active object into many and many into one. For instance, we set $N = 3$, while keep the cut-off radius at $r_c = 5$ and the wavelength $\lambda = 2$ during the simulations. Figure 4a shows the electric field distribution when there is only one line current source with a 1 A current at $z = 0$ in the conformal lens described by Eq. (13) (but with the cut-off radius introduced). The far field pattern appears to be that of three line current sources (each with a 1/3 A current) at the positions $w = e^{i\frac{\pi}{3}}, -1, e^{i\frac{5\pi}{3}}$ in vacuum, which is plotted in Figure 4b. Figure 4c plots the electric field distribution for three line current sources (each with a 1 A current) located at the positions $z = 1, e^{i\frac{2\pi}{3}}, e^{i\frac{4\pi}{3}}$ in the lens. It looks as if there is only one line current source with a 3 A current at $w = 0$ in vacuum (see in Figure 4d).

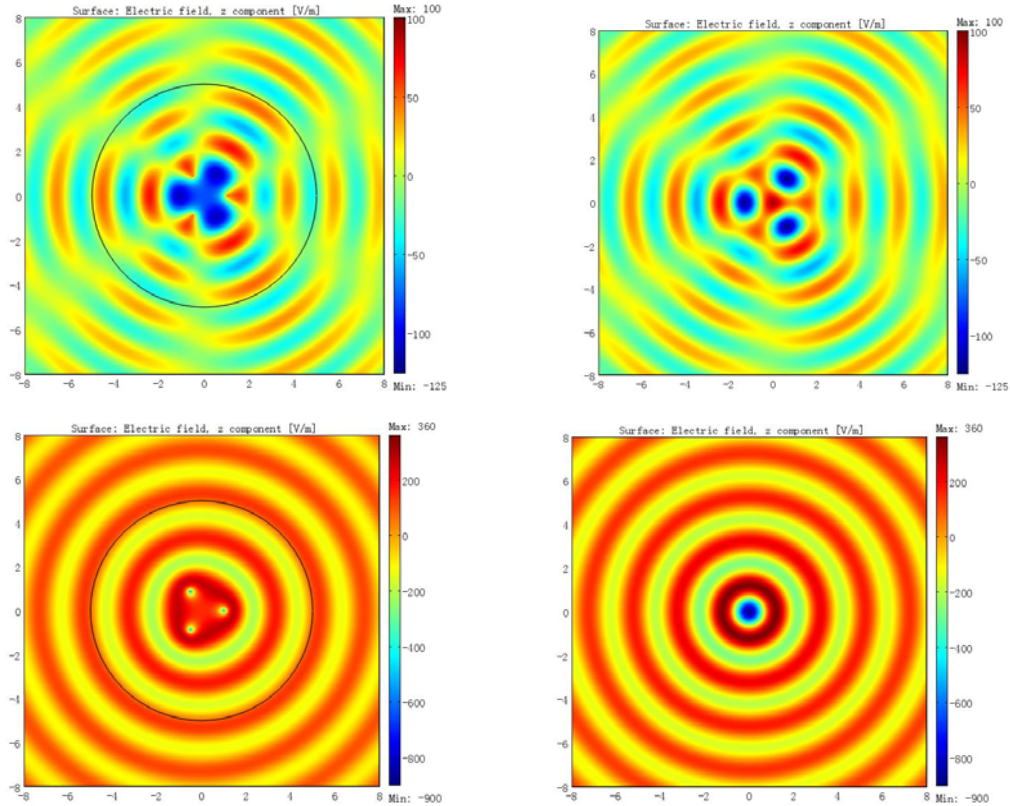

**Figure 4 | The electric field distribution for a,** one line current source with a 1 A current in the conformal lens described by Eq. (13) (yet with a cut-off radius); **b,** three line current sources (each with a 1⁄3 A current) in vacuum; **c,** three line current sources (each with a 1 A current) in the lens; **d,** one line current source with a 3 A current in vacuum.

More interesting properties of the conformal lenses for some other active objects are listed in the Supplementary Information. For example, we can transform one current sheet into two and two into one, or obtain a rotated image for one current sheet like the field rotator does [35], or transform a '+' sign into a '-' sign and vice versa.

In conclusion, we have proposed a new class of lenses to transform one active object into many and many into one using multivalued optical conformal mapping. Transmutation methods are performed to adjust the lenses for a broadband frequency range. The conformal lenses also possess some more interesting illusion properties, such as obtaining a rotated image or changing a '+' sign into a '-' sign, etc. To the best of our knowledge, it is the first time that the effect of transforming one object into many has been discovered outside of fictions [36]. For the property of transforming multiple active objects into one, applications can be further developed in enhancing brightness for the lighting systems [32]. Seeing the recent great progresses of the carpet cloaks [21, 23, 24, 25, 26], as they were derived from quasi-conformal mapping [21], which more or less resembles the optical conformal mapping [37, 38], it is not difficult to foresee the future of the optical conformal devices including the conformal lenses presented herein.

**Acknowledgements**

This work was supported by the National Natural Science Foundation of China (grant no. 11004147), the Natural Science Foundation of Jiangsu Province (grant no. BK2010211) and the Priority Academic Program Development (PAPD) of Jiangsu Higher Education Institutions.

**Supplementary Information**

In this Supplementary Information, we will introduce five intriguing properties of the conformal lenses for other active sources, such as current sheets. In the text, we mainly focused on point-to-point mappings, while here we will work on line-to-line mappings.

First of all, let us consider the two mappings below. One is that one line (from $z=-i$ to $z=i$) is mapped into two lines (from $w=-\sqrt{2}i$ to $w=-i$ and from $w=i$ to $w=\sqrt{2}i$). The other is that two lines (from $z=-\sqrt{2}$ to $z=-1$ and from $z=1$ to $z=\sqrt{2}$) are mapped into one line (from $w=-1$ to $w=1$). Again, numerical simulations will be used in describing such properties. We assign $r_c=5$, the wavelength $\lambda=2$, and $E_z=1V/m$ for each line boundary (the same applies in the following). The resembling far-field patterns in Supplementary Figure 1a and 1b show that one current sheet appears to be two, while those in 1c and 1d tell that two current sheets are transformed into one. For a much higher frequency, we can even find that the conformal lens can spilt one beam into two or combine two into one (not shown here).

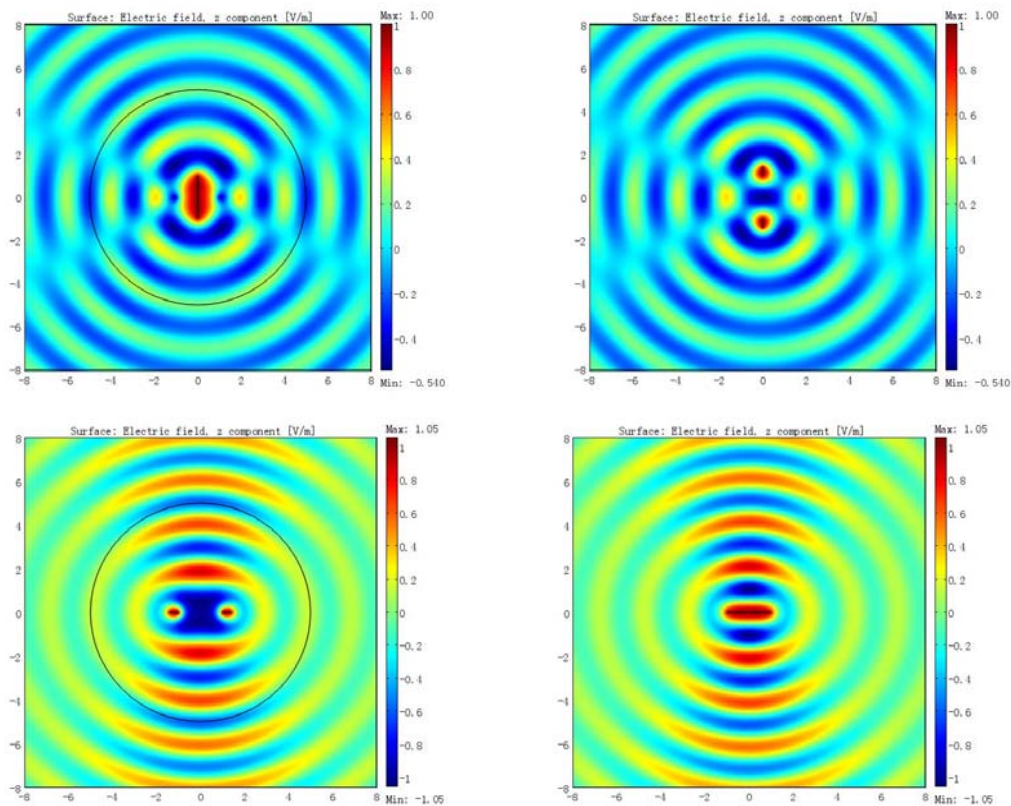

**Supplementary Figure 1 | The electric field distribution for a,** one current sheet in the conformal lens; **b,** two current sheets in vacuum; **c,** two current sheets in the lens; **d,** one current sheet in vacuum.

Next, we will consider another mapping, where a line (from $z=-1$ to $z=1$) is mapped into another (from $w=-i$ to $w=i$). Supplementary Figure 2a and 2b show almost identical far field distributions, indicating that one current sheet has a 90 deg rotation image when embedded in the conformal lens.

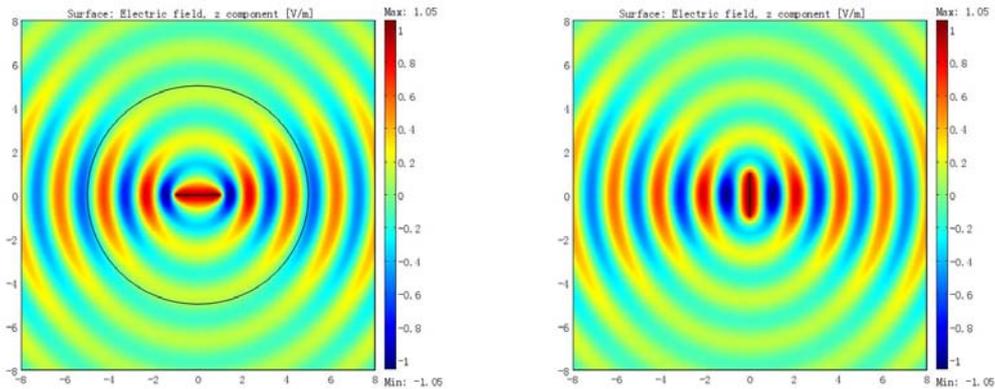

**Supplementary Figure 2 | The electric field distribution for a,** one current sheet in the conformal lens; **b,** a rotated current sheet in vacuum.

Finally, let us come to the last two interesting mappings. One is that two cross lines (one is from $z=-1$ to $z=1$, while the other is from $z=-i$ to $z=i$) are mapped into one single line (from $w=-\sqrt{2}i$ to $w=\sqrt{2}i$). The other mapping is that a line (from $z=-\sqrt{2}$ to $z=\sqrt{2}$) is mapped into two cross lines (one is from $w=-1$ to $w=1$, while the other is from $w=-i$ to $w=i$). Supplementary Figure 3a and 3b show that the conformal lens can transform two cross current sheets into one current sheet (i.e., a '+' sign into a '-' sign). Supplementary Figure 3c and 3d show that a '-' sign appears to be a '+' sign if it is embedded in the lens. These illusion effects can be treated as combinations of the above rotation effects and the effects of transforming one source into two and two into one. For example, the combination of Supplementary Figure 1a and 2a becomes 3a.

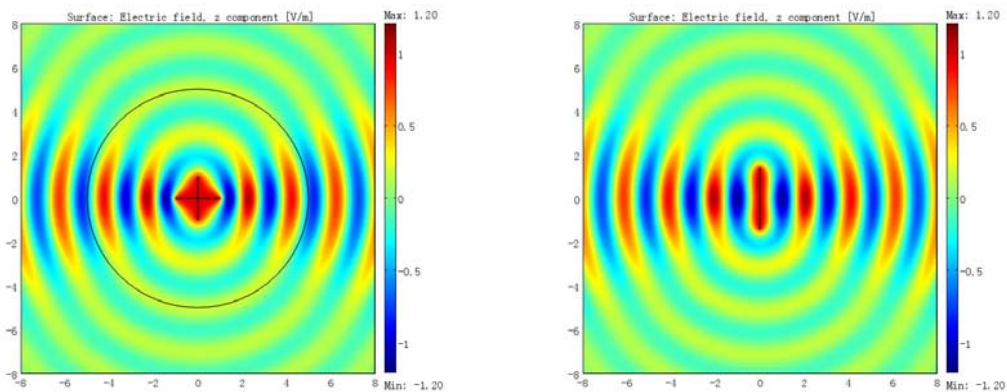

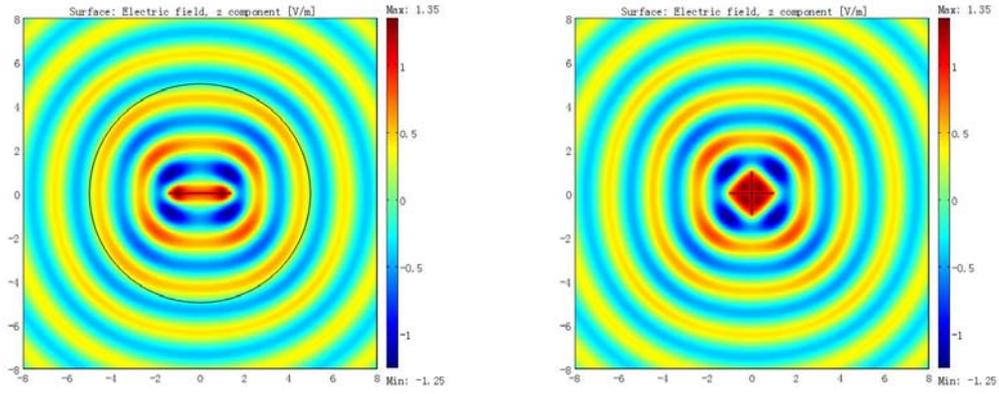

**Supplementary Figure 3 | The electric field distribution for a,** two cross current sheets in the conformal lens; **b,** one current sheet in vacuum; **c,** one current sheet in the lens; **d,** two cross current sheets in vacuum.